\documentclass[twocolumn, amsmath,amssymb,prb]{revtex4}
\usepackage{graphicx}

\begin{document}

\title{Topological Phase Transition in Layered XIn$_2$P$_2$ (X = Ca, Sr)
\footnotetext{* Corresponding author.
\\ E-mail addresses: wangtao@henannu.edu.cn }}

\author{Zhenwei Wang, Guangtao Wang$^{*}$, Xianbiao Shi, Dongyang Wang and Xin Tian}

\affiliation{College of Physics and Electronic Engineering, Henan Normal University, Xinxiang, Henan 453007, People's Republic of China}

\date{\today}

\begin{abstract}
Based on fully relativistic first-principles calculations, we studied the topological properties  of layered XIn$_2$P$_2$ (X = Ca, Sr). Band inversion can be induced by strain without SOC, forming one nodal ring in the k$_z$ = 0 plane, which is protected by the coexistence of time-reversal and  mirror-reflection symmetry. Including SOC, a substantial band gap is opened along the nodal line and the line-node semimetal would evolve into a topological insulator. These results reveal a category of materials showing quantum phase transition from trivial semiconductor and topologically nontrivial insulator by the tuneable elastic strain engineering. Our investigations provide a new perspective about the formation of topological line-node semimetal under stain. \\
\end{abstract}

\maketitle
\section{INTRODUCTION}
Topological insulator(TI)~\cite{TL1,TL2} is a new kind of material  possessing gapped bulk states and exotic metallic surface states. The robust surface state in a 3D system, sheltered against backscattering from nonmagnetic impurities as long as the bulk gap remains open and the time-reversal symmetry is preserved~\cite{serve1,serve2,serve3,serve4,serve5,serve6,serve7,serve8}. The unique feature of the surface state has attracted enormous attentions in the theoretical calculations and experimental observations~\cite{TR3,TR4}, not only because of their theoretical importance but also because of their great potential applications~\cite{app1,app2,app3}. Recently, the topological properties have been extended into a variety of three dimensional (3D) topological semimetal (TSM) systems~\cite{pro3,pro4,pro5,pro6,pro7}. In most Weyl/Dirac semimetals, conduction bands overlap with valence bands at certain momentum points. For instance, the topological properties of Na$_3$Bi~\cite{Na1,Na2} and Cd$_2$As$_3$~\cite{Ta1,Ta2} semimetals have been experimentally confirmed. However, in line-node semimetals, the band crossing points around the Fermi level (E$_F$) form a closed loop. Many systems have been proposed as line-node semimetals included Mackay-Terrones crystal (MTC)~\cite{MTC}, Bernal stacked graphene bilayer ~\cite{C}, antiperovskite Cu$_3$PdN~\cite{CuPdN} and so on. Such line-node band structure has been measured by the angle-resolved photoelectron spectroscopy (ARPES) in PbTaSe$_2$~\cite{exe2} and ZrSiS~\cite{exe3}. Interestingly, in addition to these materials, researchers  also found node line structures in the photonics crystals~\cite{pho} and spin liquids~\cite{liq}. The line-node material is becoming a hot topic.

The topological line-node semimetal could be driven into 3D TI or Dirac semimetal by the SOC and strain. For example, CaTe~\cite{Ca} is a TSM, possessing the nodal rings without SOC, exhibiting Dirac semimetal behavior when SOC is included. When it was applied by appropriate strain, it becomes a strong topological insulator~\cite{Ca}. However, the CaAgX (X = P, As) can  be driven into a Tl phase from line-node semimetal by taking into account of SOC~\cite{Ag}.

Bulk CaIn$_2$P$_2$ and SrIn$_2$P$_2$ are layered semiconductors with indirect-gap and direct-gap respectively. They have been investigated with their electronic and optical properties~\cite{opt}.  Up to now, their topological properties have not been studied.  In this study, we find that the indirect-gap semiconductor can be changed into `direct'-gap by applying uniaxial strain slightly. Further Increasing strain, the compound undergoes a transition from trivial insulator to a line-node semimetal with a nodal ring in the k$_z$ = 0 plane. When SOC is taken into account, the compounds become a nontrivial topological insulator with Z$_2$ index (1,000). Our results provide a new perspective to understand the effect of strain  and SOC on the formation of topological line-node semimetal and insulator.

\section{CRYSTAL STRUCTURE AND METHOD}

The ternary compound CaIn$_2$P$_2$ is isostructural with SrIn$_2$P$_2$ and crystallizes in a hexagonal structure with the space group P63/mmc~\cite{struct}. The alkaline earth cations X (Ca or Sr) are located at a site with $\overline{3}$m symmetry; In and P are located at sites with 3m symmetry. The lattice constants of CaIn$_2$P$_2$ and SrIn$_2$P$_2$ are a=b= 4.022 \AA, c= 17.408 \AA and a=b= 4.094 \AA, c= 17.812 \AA, respectively~\cite{exe2}. The corresponding unit cell of the XIn$_2$P$_2$ (X = Ca, Sr) compounds is depicted in Fig.1(a). It contains two chemical formula units, the layers of X$^{2+}$ cations are separated by [In$_2$P$_2$]$^{2-}$ layers. The atoms  positioned at the following Wyckoff positions: X (Ca, Sr): 2a (0, 0, 1/2), In: 4f (2/3, 1/3, 0.329) and P: 4f (1/3, 2/3, 0.396). The bulk Brillouin zone (BZ) and projected surface Brillouin zones of (001) plane are illustrated in Fig.1(b).
\begin{figure}[htbp]
\begin{center}
\includegraphics[clip,scale=0.16]{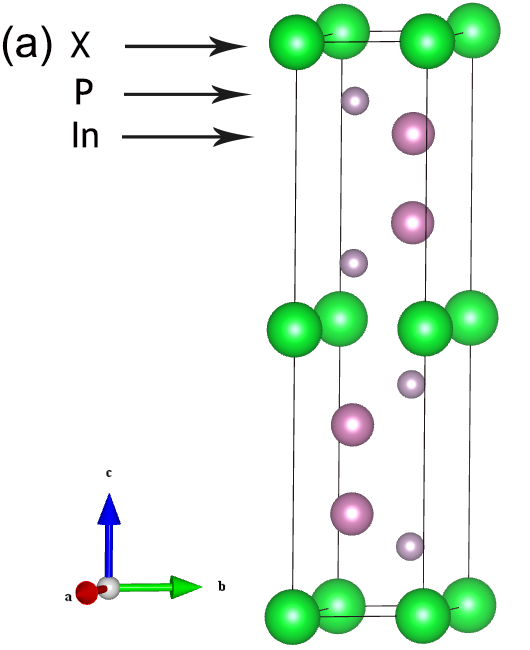}
\includegraphics[clip,scale=0.32]{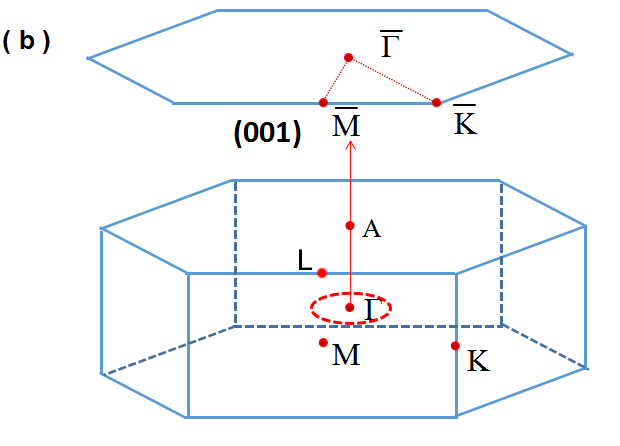}
\caption{(Color online) (a) The Crystal structure of XIn$_2$P$_2$ with P63/mmc symmetry. (b) Brillouin zone of bulk and the projected surface Brillouin zones of (001) plane, as well as high-symmetry points. There is a line-node ring lies in the k$_z$=0 plane.}
\end{center}
\end{figure}

Firstly, we perform density functional calculations by using the WIEN2K package~\cite{WIEN},  with the modified Becke-Johnson exchange potential (mBJ) to get the accurate band gap~\cite{mbj}. The valence configurations of the Ca, Sr, P and In atoms  are  as Ca $3s^23p^64s^2$, Sr $4s^24p^65s^2$, P $3s^23p^3$ and In $4d^{10}5s^25p^1$, respectively. The plane-wave cutoff parameter R$_{MT}$K$_{max}$ is set to 7 and a 12$\times$ 12$\times$ 3 K-mesh is used for the BZ integral. The SOC interaction is included by using the second-order variational procedure.  The tight binding model based on maximally localized Wannier functions (MLWF) method~\cite{MLWF1,MLWF2} has been constructed in order to investigate the surface states. The surface Green's function of the semi-infinite system, whose imaginary part is the local density of states to obtain the dispersion of the surface states, can be calculated through an iterative method\cite{sancho1984,sancho1985,wuquansheng}.  To conform the compounds are  dynamicly stable, the phonon spectrum was calculated by using  the PHONON code~\cite{ph}.

\section{Results and discussion}

The electronic band structures of the unstrained  and 5\% strained CaIn$_2$P$_2$ (a/a$_0$=1.05) with SOC are depicted in Fig.2(a) and (b). The unstrained  CaIn$_2$P$_2$ has an indirect band gap about 0.5 eV, which agrees with previous studies~\cite{opt}. In Fig.2(a), the $\Gamma$8$^+$  band was pushed down, while the $\Gamma$8$^-$ was pushed up, when we induced tensile stress in the a-b plane and   compressive stress along c-axis, i.e. increasing a-axis and  decreasing c-axis with fixed cell volume. When a$\geqslant$1\%a$_0$, the compound becomes direct band gap insulator. At the critical point a=2\%a$_0$, the $\Gamma$8$^+$ and $\Gamma$8$^-$ touch each other. When we further increase a-axis to a=5\%a$_0$ the $\Gamma$8$^+$ dropped below $\Gamma$8$^-$ band in Fig.2(b). Comparing Fig.2(a) and Fig.2(b), we find that it is the strain rather than the SOC that induces the `band-invertion'~\cite{PRL108}. From the inset figure in Fig.2(b), we can see a band gap about 35 meV. It is well know that the inversion of bands with opposite parity is a strong indication of the formation of topologically nontrivial phases. This suggests that the strain induces a topological phase transition in CaIn$_2$P$_2$. The band structure of SrIn$_2$P$_2$ has the similar property, excepted its 0.3 eV direct bang gap.

 \begin{figure}[htbp]
\includegraphics[clip,scale=0.21]{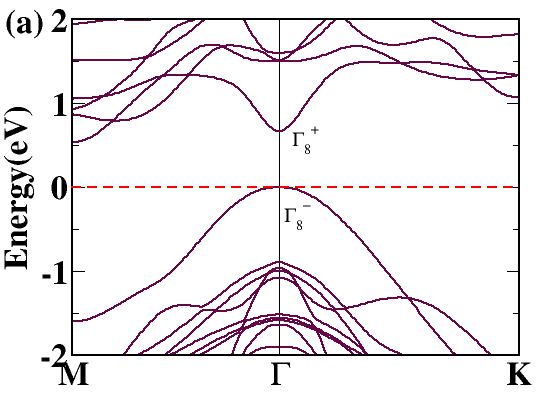}
\includegraphics[clip,scale=0.21]{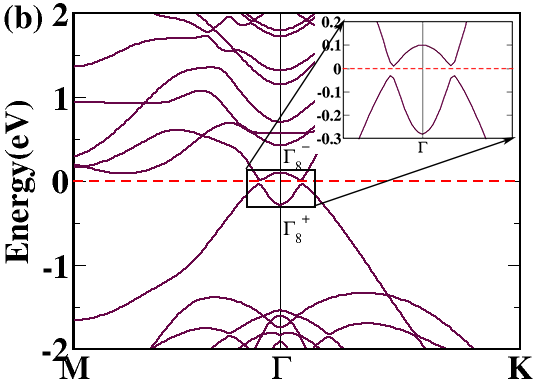}
\caption{(Color online)  Electronic band structures of unstrained CaIn$_2$P$_2$  with SOC. (b) Electronic band structures of band-inverted CaIn$_2$P$_2$ (a/a$_0$=1.05) with SOC. The details of band-inverted CaIn$_2$P$_2$ around E$_F$ are shown in the inset.}
\end{figure}

\begin{figure}[htbp]
\includegraphics[clip,scale=0.3]{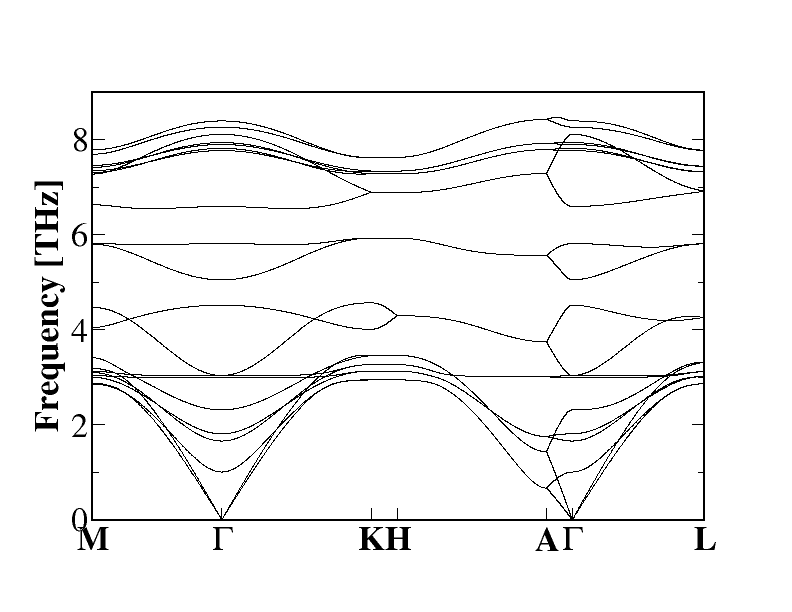}
\caption{ Phonon dispersion in SrIn$_2$P$_2$ along the high-symmetry directions.}
\end{figure}
To confirm the topological nature of CaIn$_2$P$_2$, all four Z$_2$ topological invariants (v0; v1 v2 v3 )~\cite{Fu} were calculated before ( zero strained ) and after ( 5\% strained ) the band inversion structures. Following the  method developed by Fu and Kane~\cite{Fu}, we calculate  directly the  Z$_2$ topological invariants from knowledge about the parity of each pair of Kramer's degenerate occupied energy bands at the eight time-reversal momenta (1$\Gamma$, 3M, 1A, and 3L), because of the existence of inversion symmetry  in our compounds. Before the band inversion, the products of parities  $\delta$$_i$  are identical, yielding a value of zero for all four Z$_2$ topological invariants, i.e., a topologically trivial state (0;000). When the bands with $\Gamma$8$^+$ and $\Gamma$8$^-$ symmetries switch around E$_F$, see Figs. 2(a) and 2(b), $\delta$$_\Gamma$ changes its sign while all other seven $\delta$$_i$ remain unchanged. The parities of time-reversal momenta are listed in Table I. It gives rise to a topological state with Z$_2$ class (1;000) and confirms that the strain induces a topological phase transition.

Generally, the band inversion is generated by SOC, formatting the topologically nontrivial state, similar to the prototypical 3D topological insulator Sr$_2$Pb and Bi$_2$Te$_3$~\cite{fan1,fan2}. However, in our study, the band inversion was induced by the strain only, without considering any SOC. Then, what roles the strain and SOC play in the topological phase transformation of XIn$_2$P$_2$? In Fig.4, we take SrIn$_2$P$_2$ as the example to answer the question. Fig.4 shows the orbital characteristic band structure ( around the $\Gamma$ and near Fermi energy ) of SrIn$_2$P$_2$ under the different strains. To quantify the effects of strain and SOC on the band inversion in SrIn$_2$P$_2$, we temporally switch off the SOC in Figs.4(a) and 4(b). For the unstrained compound ( Figs.4a ), there is a direct band gap about 0.3 eV,  with the In-s state  above the P-p state, indicating a trivial topological insulator. When we induce the strain with a=5\%a$_0$ and without SOC, the In-s state was pushed down the P-p state, resulting band inversion and line-node band structure ( Fig.4b ). This result indicates that the  band inversion is caused by the strain rather than the SOC. For the 5\% strained compound, when we include SOC, there is a band gap about 35 meV exactly at the Fermi energy. Comparing the three figures in Fig.4, we can draw the conclusion that lattice strain induces the `band inversion' while SOC opens the band gap.

\begin{table}[tbp]
\renewcommand\arraystretch{1.15}
\begin{center}
\caption{The parities of time-reversal invariant k points with SOC and their corresponding topological invariant.}
\begin{tabular}{lcccc|ccccccccccccccccccccccccc}
\hline\hline
&\multicolumn{1}{c}{} &\multicolumn{1}{c}{} &\multicolumn{1}{c}{} &\multicolumn{1}{c}{} &\multicolumn{1}{c}{$\Gamma$} &\multicolumn{1}{c}{} &\multicolumn{1}{c}{} &\multicolumn{1}{c}{} &\multicolumn{1}{c}{}&\multicolumn{1}{c}{} &\multicolumn{1}{c}{M} &\multicolumn{1}{c}{} &\multicolumn{1}{c}{} &\multicolumn{1}{c}{} &\multicolumn{1}{c}{}&\multicolumn{1}{c}{}&\multicolumn{1}{c}{A} &\multicolumn{1}{c}{} &\multicolumn{1}{c}{} &\multicolumn{1}{c}{}&\multicolumn{1}{c}{}&\multicolumn{1}{c}{} &\multicolumn{1}{c}{L} &\multicolumn{1}{c}{} &\multicolumn{1}{c}{}&\multicolumn{1}{c}{}&\multicolumn{1}{c}{} &\multicolumn{1}{c}{} &\multicolumn{1}{c}{Z$_{2}$} \\
\hline
SOC, strain=0$\%$& & & &  &+ & & & & &  &+ & & & & &    &+  & & & & &   &+ & & & & &   & 0 \\
SOC, strain=5$\%$& & & &  &- & & & & &  &+  & & & & &   &+  & & & & &   &+ & & & & &   & 1 \\
\hline\hline
\end{tabular}
\end{center}
\end{table}

To gain further insight into the mechanism of the strain induced band inversion, we study the electronic structure of unstrained SrIn$_2$P$_2$ around E$_F$ in more detail. Without SOC, the projected band structure of SrIn$_2$P$_2$ has been shown in Fig.5(a). It shows that the valence bands $\Gamma$$_6$ and $\Gamma$$_4$ are dominated by the P-p$_z$ states, while the conduction bands $\Gamma$$_2$ and $\Gamma$$_3$ are mainly composed of In-s state.  Both the valence bands  and conduct bands are doubly degenerated at A-point, and they split up into $\Gamma$$_{2}$$^-$/$\Gamma$$_{3}$$^+$ and $\Gamma$$_{4}$$^-$/$\Gamma$$_{6}$$^-$ bands, respectively.  As a result, the $\Gamma$$_{3}$$^+$ and $\Gamma$$_{4}$$^-$ bands give rise to the conduction band minimum and valence band maximum, respectively. The relative position of the $\Gamma$$_{3}$$^+$ and $\Gamma$$_{4}$$^-$ bands decides whether a band inversion occurs or not. The energy difference $\Delta$E(A)  is determined by the intra-slab covalent s-p$_z$ hybridization, while $\Delta$E($\Gamma$$_{4,6}$) and $\Delta$E($\Gamma$$_{2,3}$) are related to inter-slab van der Waals p$_z$-p$_z$ and s-s interactions. This observation is in line with findings for layered transition metal dichalcogenide semiconductors, such as MoS$_2$~\cite{MoS1,MoS2}. We have determined the four parameters $\Delta$E(A), $\Delta$E($\Gamma$$_{4,6}$), $\Delta$E($\Gamma$$_{2,3}$), and $\Delta$E$_g$ as functions of the uniaxial strain (a-a$_0$)/a$_0$ (volume is fixed). The results are presented in Fig.5(b). With increasing tensile strain in the ab-plane and compressive strain along c-axis, the intra-slab interaction decreases, resulting the increment of the energy spilt $\Delta$E($\Gamma$$_{4,6}$). However, the inter-slab interaction increase with c-axis decreasing, resulting a strongly reduced $\Delta$E(A), indicating strongly intra-slab covalent s-p$_z$ hybridization. The energy level $E_{\Gamma_4}$ ( derived from P-p$_z$ orbit ) increases with c-axis decreasing. So the band gap E$_g$ decreases with the strain increment, and the band gap closes at the critical point $\delta$=0.02. With further increasing the strain, the band inversion takes place and the compound becomes topological nontrivial line-node semimetal. SOC opens the band gap along the nodal ring and drives the compound into topological nontrivial insulator.

\begin{figure}[tbp]
\includegraphics[clip,scale=0.17]{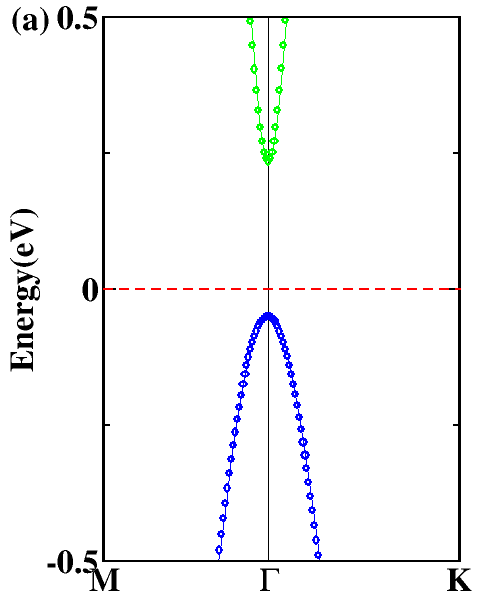}
\includegraphics[clip,scale=0.17]{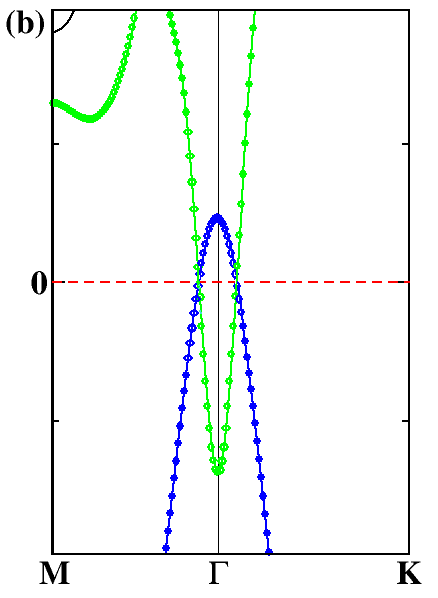}
\includegraphics[clip,scale=0.17]{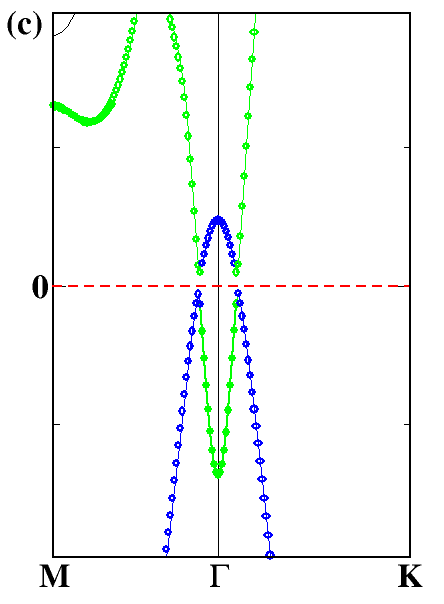}
\caption{(Color online) The orbital characteristic band structures (  along the M - $\Gamma$ - K direction, near Fermi energy )  of SrIn$_2$P$_2$ under the different strains. (a) zero strain without SOC, (b) 5\% of uniaxial strain without SOC, (c) 5\% of uniaxial strain with SOC. The weight of atomic orbital In(s) and P(p) is proportional to the radius of the green (blue) circle. Band inversion could be seen clearly around the $\Gamma$ point.}
\end{figure}

\begin{figure}[htbp]
\begin{center}
\includegraphics[clip,scale=0.205]{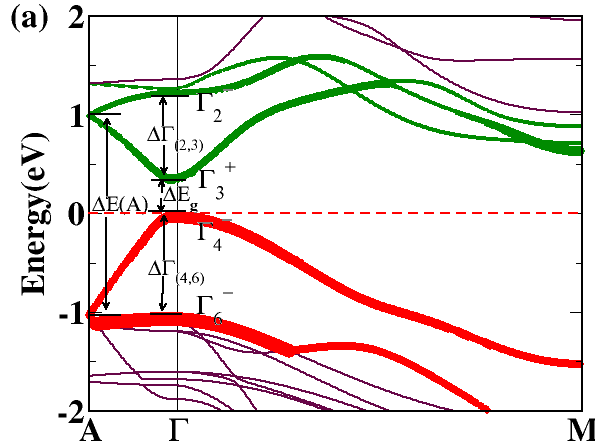}
\includegraphics[clip,scale=0.17]{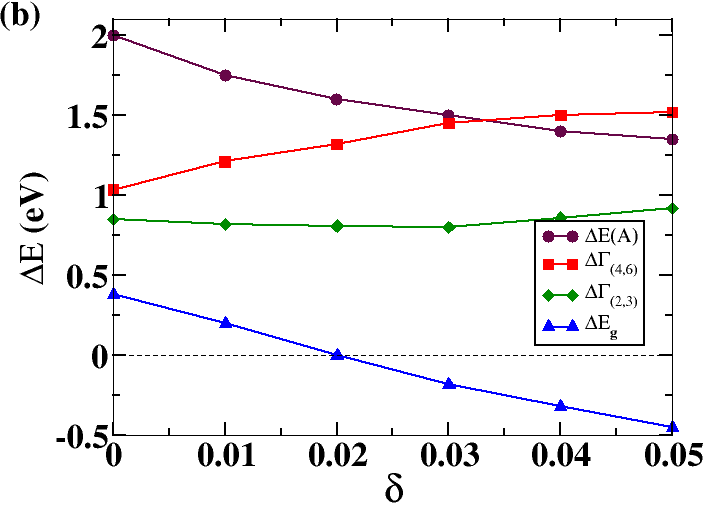}
\caption{(Color online)(a) The band structure of unstrained SrIn$_2$P$_2$ without SOC, weighted with the In-s and P-P$_z$ characters. In-s and P-P$_z$ states are distinguished by red and green colors. (b) $\Delta$E(A), $\Delta$E($\Gamma$$_{4,6}$), $\Delta$E($\Gamma$$_{2,3}$) and $\Delta$E$_g$ as functions of the inter-slab tensile strain $\delta$=(a-a$_0$)/a$_0$ (volume is fixed), respectively.}
\end{center}
\end{figure}

\begin{figure*}[htbp]
\begin{center}
\includegraphics[clip,scale=0.163]{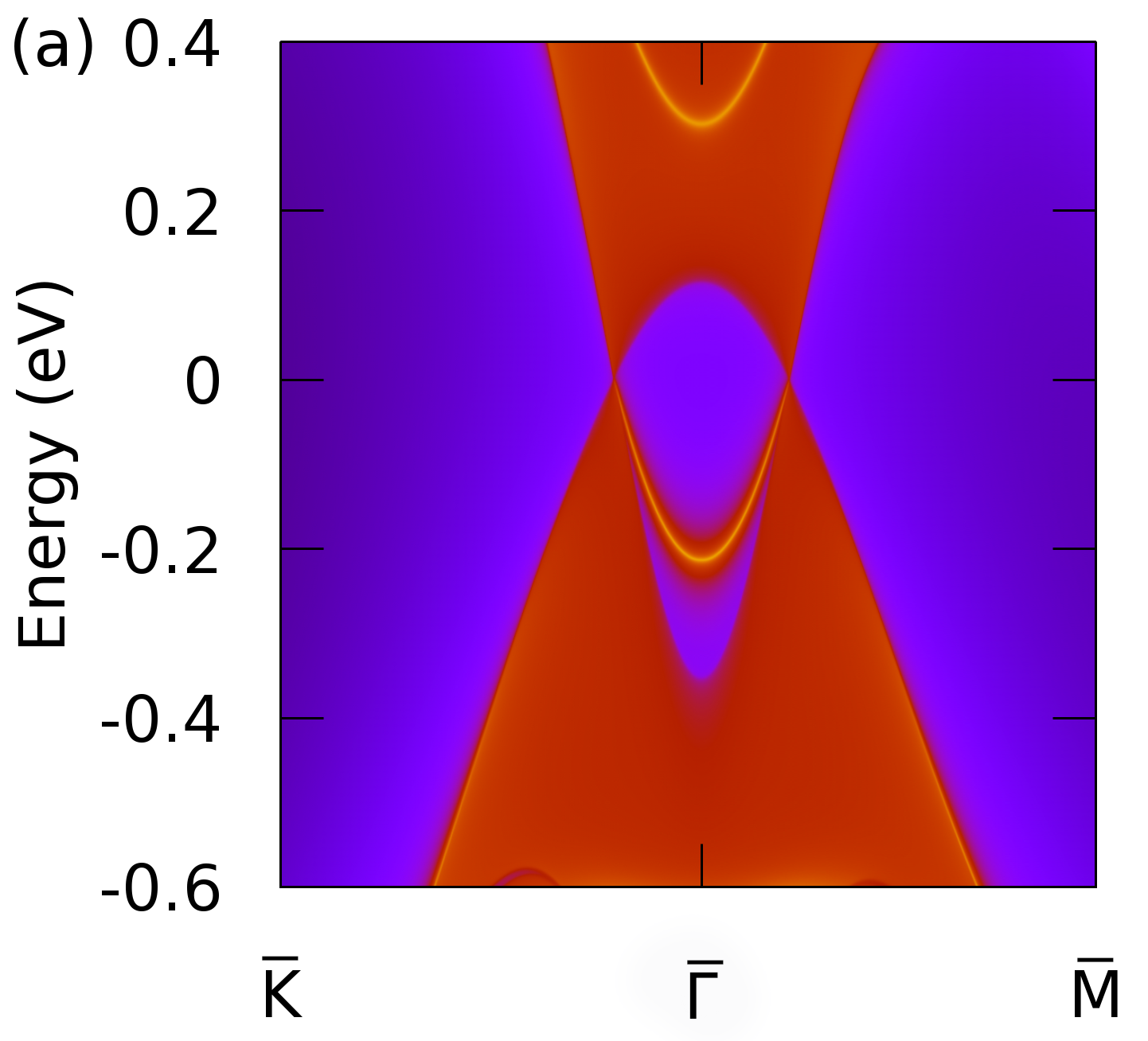}\vspace{0.1cm}
\includegraphics[clip,scale=0.163]{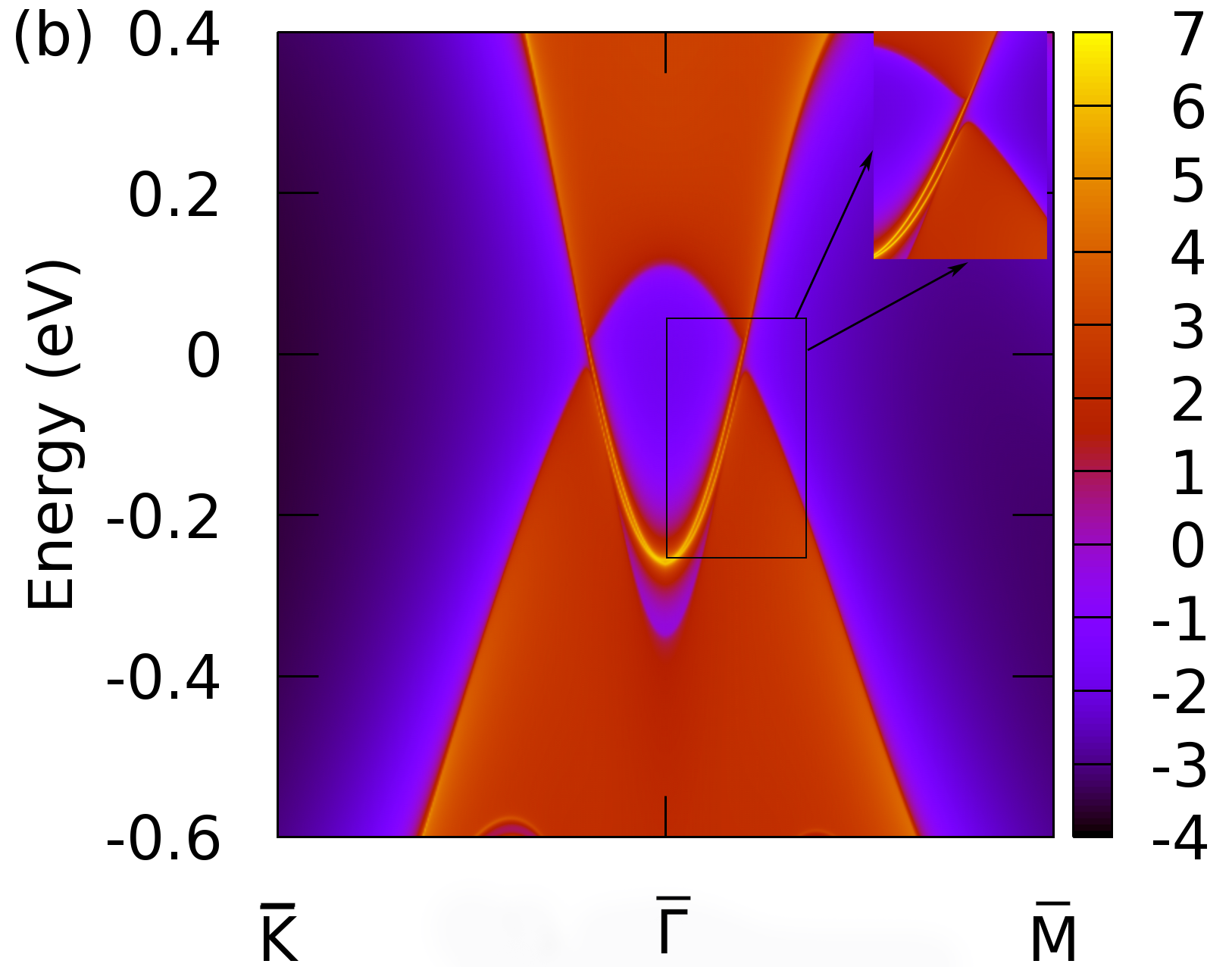}\vspace{0.1cm}
\caption{(Color online) Surface density of states (SDOS) in SrIn$_2$P$_2$ (a) without and (b) with SOC. The yellow lines highlight the topologically protected metallic surface states for the (001) surface.}
\end{center}
\end{figure*}


For the topological materials, the existence of the topological prevent surface states is an important character. So we calculate  the (001) surface of SrIn$_2$P$_2$ by preforming the WANNIER$\_$TOOLS package~\cite{tool} in a tight-binding (TB) scheme based on the maximally localized Wannier functions (MLWFs)~\cite{MLWF1}. The surface state of Sr$_2$In$_2$P$_2$ (001) surface without and with SOC have been shown in Fig.6(a) and (b), respectively. When SOC is ignored, the system is a line-node semimetal, with the bulk bands touched along $\bar{K}-\bar{\Gamma}$, $\bar{\Gamma}-\bar{M}$ lines and a surface state lining two touching points. This is  similar to CaAgAs~\cite{Ag}, protected by the coexistence of time-reversal and mirror-reflection symmetry. When SOC is included, a gap opened at the two touching points, and the compound became an topological nontrivial insulator, with two surfaces states connecting bulk valance and conduction bands (see Fig.6b), coinciding with the previous calculated  Z$_2$ ( Table. I).

\section{CONCLUSION}

In conclusion, by the first-principles calculations,  we find the quantum phase transition in XIn$_2$P$_2$ from conventional semiconductor  into the  line node semimetal by strain engineering only without  SOC. While SOC takes the role as opening band gap along the line node and gives rise to the nontrivial topological insulator state.

\begin{acknowledgments}
The authors acknowledge support from the NSF of China (No.11274095, No.10947001) and the Program for Science and Technology Innovation Talents in the Universities of Henan Province (No.2012HASTIT009, No.104200510014, and No.114100510021). This work is supported by The High Performance Computing Center (HPC) of Henan Normal University.
\end{acknowledgments}


\begin{thebibliography}{}
\bibitem{TL1}J. Moore, Nat. Phys. 5, 378 (2009).
\bibitem{TL2}M. Z. Hasan and C. L. Kane, Rev. Mod. Phys. 82, 3045 (2010).
\bibitem{serve1}M. Z. Hasan and C. L. Kane, Rev. Mod. Phys. 82, 3045 (2010).
\bibitem{serve2}X. L. Qi and S. C. Zhang, Rev. Mod. Phys. 83, 1057 (2011).
\bibitem{serve3}B. A. Bernevig, T. L. Hughes, and S.-C. Zhang, Science 314, 1757 (2006).
\bibitem{serve4}C. L. Kane and E. J. Mele, Phys. Rev. Lett. 95, 226801 (2005).
\bibitem{serve5}H.Zhang,C.-X.Liu,X.-L.Qi,X.Dai,Z.Fang,andS.-C.Zhang, Nat. Phys. 5, 438 (2009).
\bibitem{serve6}K?nig M, Wiedmann S, Brš¹ne C, et al, Science 318, 766 (2007).
\bibitem{serve7}J. Moore, Nat. Phys. 5, 378 (2009).
\bibitem{serve8}X.-L. Qi and S.-C. Zhang, Phys. Today 63, 33 (2010).
\bibitem{TR3}X.-L. Qi and S.-C. Zhang, Phys. Today 63, 33 (2010).
\bibitem{TR4}X.-L. Qi and S.-C. Zhang, Rev. Mod. Phys. 83, 1057 (2011).
\bibitem{app1}Fu, Liang, and Charles L. Kane, Phys. Rev. Lett. 100, 096407 (2008).
\bibitem{app2}Qi X L, Hughes T L and Zhang S C Phys. Rev. B, 78, 195424 (2008). Qi X-L, Li R, Zang J and Zhang S-C, Science 323, 1184 (2009).
\bibitem{app3}Essin A M, Moore J E and Vanderbilt D, Phys. Rev. Lett. 102, 146805 (2009).
\bibitem{pro3}X. Wan, A. M. Turner, A. Vishwanath, and S. Y. Savrasov, Phys. Rev. B 83, 205101 (2011).
\bibitem{pro4}L. Balents, Physics 4, 36 (2011).
\bibitem{pro5}Y. Ando, J. Phys. Soc. of Jpn. 82, 102001 (2013).
\bibitem{pro6}K.-Y. Yang, Y.-M. Lu, and Y. Ran, Phys. Rev. B 84, 075129 (2011).
\bibitem{pro7}T. O. Wehling, A. M. Black-Schaffer, and A. V. Balatsky, Adv. Phys. 63, 1 (2014).
\bibitem{Na1}Z. Wang, Y. Sun, X.-Q. Chen, C. Franchini, G. Xu, H. Weng, X. Dai, and Z. Fang, Phys. Rev. B 85, 195320 (2012).
\bibitem{Na2}Z. K. Liu, B. Zhou, Y. Zhang, Z. J. Wang, H. M. Weng, D. Prabhakaran,S.-K.Mo, Z.X.Shen, Z.Fang, X.Dai, Z.Hussain and Y. L. Chen, Science 343, 864 (2014).
\bibitem{Ta1}Z. Wang, H. Weng, Q. Wu, X. Dai, and Z. Fang, Phys. Rev. B 88, 125427 (2013).
\bibitem{Ta2}Z. K. Liu, J. Jiang, B. Zhou, Z. J. Wang, Y. Zhang, H. M. Weng, D. Prabhakaran, S.-K. Mo, H. Peng, P. Dudin, T. Kim, M. Hoesch, Z. Fang, X. Dai, Z. X. Shen, D. L. Feng, Z. Hussain, and Y. L. Chen, Nat. Mater. 13, 677 (2014).
\bibitem{MTC}Z. Gao, M. Hua, H. Zhang, and X. Zhang, arXiv:1507.07504.
\bibitem{C}J.-W. Rhim and Y. B. Kim, Phys. Rev. B 92, 045126 (2015).
\bibitem{CuPdN}Yu R, Weng H, Fang Z, et al. Physical review letters, 2015, 115(3): 036807.
\bibitem{exe2}G. Bian, T.-R. Chang, R. Sankar, S.-Y. Xu, H. Zheng, T. Neupert, C.-K. Chiu, S.-M. Huang, G. Chang, I. Belopolski, D. S. Sanchez, M. Neupane, N. Alidoust, C. Liu, B. Wang, C.-C. Lee, H.-T. Jeng, A. Bansil, F. Chou, H. Lin, and M. Zahid Hasan, arXiv:1505.03069.
\bibitem{exe3}L. M. Schoop, M. N. Ali, C. Straer, V. Duppel, S. S. P. Parkin, B. V. Lotsch, and C. R. Ast, arXiv:1509.00861.
\bibitem{pho}L. Lu, L. Fu, J. D. Joannopoulos, and M. Soljai, Nat. Photonics 7, 294 (2013).
\bibitem{liq}W. M. H. Natori, E. Miranda, and R. G. Pereira, arXiv:1505.06171.
\bibitem{Ca}Du Y, Tang F, Wang D, et al. arXiv preprint arXiv:1605.07998, 2016.
\bibitem{Ag}Yamakage A, Yamakawa Y, Tanaka Y, et al. Journal of the Physical Society of Japan, 85(1): 013708 (2015).
\bibitem{opt}Guechi N, Bouhemadou A, Guechi A, et al. Journal of Alloys and Compounds, 577: 587-599 (2013).
\bibitem{struct}J.F. Rauscher, C.L. Condron, T. Beault, S.M. Kauzlarich, N. Jensen, P. Klavins, S. MaQuilon, Z. Fisk, M.M. Olmstead, Acta Cryst. C 65, 69 (2009).
\bibitem{WIEN}P. Blaha, K. Schwarz, G.K.H. Madsen, D. Kvasnicka, and L. Luitz, (Technical University of Vienna, Vienna, 2001).
\bibitem{mbj}F. Tran and P. Blaha, Phys. Rev. Lett. 102, 226401 (2009).
\bibitem{MLWF1}A. A. Mostofi, J. R. Yates, Y.-S. Lee, I. Souza, D. Vanderbilt, and N. Marzari, Comput. Phys. Commun. 178, 685 (2008).
\bibitem{MLWF2}N. Marzari, A. A. Mostofi, J. R. Yates, I. Souza, and D. Vanderbilt, Rev. Mod. Phys. 84, 1419 (2012).
\bibitem{sancho1984} J. M .Sancho, J. Math. Phys., 25(2), 354 (1984).
\bibitem{sancho1985} J. M. Sancho, Phys. Rev. A 31 (5), 3523 (1985 ).
\bibitem{wuquansheng} G. W. Winkler, Q. S. Wu, et. al,, arxiv:1602.07001; A. Tamai, Q. S. Wu, I. Cucchi, and et. al., Phys. Rev. X 6, 031021 (2016).
\bibitem{ph}X. Gonze and C. Lee, Phys. Rev. B 55, 10355 (1997).
\bibitem{PRL108}Zhu Z, Cheng Y, Schwingenschl?gl U, Physical review letters, 108(26): 266805 (2012).
\bibitem{Fu}L. Fu and C.L. Kane, Phys. Rev. B 76, 045302 (2007).
\bibitem{fan1}H.Zhang, C.-X.Liu, X.-L.Qi, X.Dai, Z.Fang,andS.-C.Zhang, Nat. Phys. 5, 438 (2009).
\bibitem{fan2}Y. Sun, X.-Q. Chen, C. Franchini, D. Li, S. Yunoki, Y. Li, and Z. Fang, Phys. Rev. B 84, 165127 (2011).
\bibitem{MoS1}Zhu, Z. Y., Y. C. Cheng, and Udo Schwingenschl?gl., Phys. Rev. B 84, 153402 (2011).
\bibitem{MoS2}K.F. Mak, C. Lee, J. Hone, J. Shan, and T.F. Heinz, Phys. Rev. Lett. 105, 136805 (2010).
\bibitem{tool}Q. S. Wu, https://github.com/quanshengwu/wannier$\_$tools (2015).
\bibitem{super}N. Kopnin, T. Heikkil, and G. Volovik, Phys. Rev. B 83, 220503 (2011).

\end{thebibliography}
\end{document}